\documentclass[final,5p,times,twocolumn]{elsarticle}
\pdfoutput=1

\usepackage{graphicx}

\usepackage{amssymb}

\usepackage{bm}
\usepackage{latexsym}

\journal{arXiv}

\makeatletter 
\DeclareRobustCommand\onlinecite{\@onlinecite} 
\def\@onlinecite#1{\begingroup\let\@cite\NAT@citenum\citealp{#1}\endgroup} 
\makeatother

 \newcommand{\iotabar}{\mbox{$\,\iota\!\!$-}}
 \newcommand{\Fig}[1]{Fig.~\ref{fig:#1}}
 
 \newcommand{\Eqn}[1]{Eq.\,(\ref{eq:#1})}
 \newcommand{\Sec}[1]{Sec.\,\ref{sec:#1}}

 \renewcommand{\vec}[1]{\mbox{$\bf #1$}}
 \newcommand{\const}{\mbox{const}}
 \newcommand{\grad}{\mbox{\boldmath\(\nabla\)}}
 
 \newcommand{\curl}{  \mbox{\boldmath\(\nabla\times\)} }
 \newcommand{\dotv}{  \mbox{\boldmath\(\cdot\)} }
 \newcommand{\cross}{  \mbox{\boldmath\(\times\)} }
 \newcommand{\esub}[1]{ {\bf e}_{#1} }
 \newcommand{\esup}[1]{ {\bf e}^{#1} }
 \newcommand{\dsub}[1]{ \partial_{#1} }
 \newcommand{\Id}{\textsf{\textbf{I}}} 
 \newcommand{\Proj}{\textsf{\textbf{P}}} 
 \newcommand{\be}{\begin{eqnarray}}
 \newcommand{\ee}{\end{eqnarray}}
 
 \newcommand{\Poincare}{Poincar{\'e}}

\begin{document}

\begin{frontmatter}



\title{Are ghost surfaces quadratic-flux-minimizing?}
\author{S.R. Hudson}
\ead{shudson@pppl.gov}
\address{Princeton Plasma Physics Laboratory, PO Box 451, Princeton NJ 08543, USA}
\author{R.L. Dewar}
\address{Plasma Research Laboratories, Research School of Physics \& Engineering, The Australian National University, Canberra, ACT 0200, Australia}
\ead{robert.dewar@anu.edu.au}
\date{Draft Version PLA1.1, 6th Sep. 2009}


\begin{abstract}
Two candidates for ``almost-invariant'' toroidal surfaces passing through magnetic islands, namely quadratic-flux-minimizing (QFMin) surfaces and ghost surfaces, use families of periodic pseudo-orbits (i.e. paths for which the action is not exactly extremal). 
QFMin pseudo-orbits, which are coordinate-dependent, are field lines obtained from a modified magnetic field, and ghost-surface pseudo-orbits are obtained by displacing closed field lines in the direction of steepest descent of magnetic action, $\oint \vec{A}\dotv\vec{dl}$. 
A generalized Hamiltonian definition of ghost surfaces is given and specialized to the usual Lagrangian definition.  
A modified Hamilton's Principle is introduced that allows the use of Lagrangian integration for calculation of the QFMin pseudo-orbits. 
Numerical calculations show QFMin and Lagrangian ghost surfaces give very similar results for a chaotic magnetic field perturbed from an integrable case, and this is explained using a perturbative construction of an auxiliary poloidal angle for which QFMin and Lagrangian ghost surfaces are the same up to second order. 
While presented in the context of 3-dimensional magnetic field line systems, the concepts are applicable to defining almost-invariant tori in other $1\frac{1}{2}$ degree-of-freedom nonintegrable Lagrangian/Hamiltonian systems.
\end{abstract}

\begin{keyword}
Toroidal magnetic fields \sep Hamiltonian dynamics \sep Lagrangian dynamics \sep almost-invariant tori
\PACS 05.45.-a \sep 52.55.Dy \sep 52.55.Hc
\end{keyword}

\end{frontmatter}

\section{introduction}

 The understanding of nonintegrable Hamiltonian systems  is greatly simplified if one can construct a coordinate framework based on a set of surfaces that are either invariant under the dynamics or, where this is impossible, surfaces that are {\em almost}-invariant.  
As invariant tori and cantori in nonintegrable systems can be approximated by sequences of periodic orbits, the theory of almost-invariant surfaces is built around periodic orbits, which constitute the remanent invariant sets surviving after integrability is destroyed by symmetry-breaking perturbations.
We consider two classes of almost-invariant surfaces, \emph{quadratic-flux-minimizing} (QFMin) surfaces \cite{Dewar_Hudson_Price_94} and \emph{ghost surfaces} \cite{Hudson_Dewar_96,Gole_01}. 

Almost-invariant tori are important in the theory of magnetic confinement of toroidal plasmas, in particular to the theory of transport in chaotic magnetic fields \cite{Hudson_Breslau_08}, and we set this paper in the context of the nonintegrable magnetic fields, $\vec{B}$, encountered in devices without a continuous symmetry.
  However, as magnetic field lines are orbits of a $1\frac{1}{2}$ degree-of-freedom Hamiltonian system, \cite{Flux_Coords} the discussion is applicable, with appropriate translations of terminology, to any such system---e.g. in this paper we use ``magnetic field line'' and ``orbit'' interchangeably.  

\begin{figure}[htbp]
	\centering\includegraphics[width=3in]{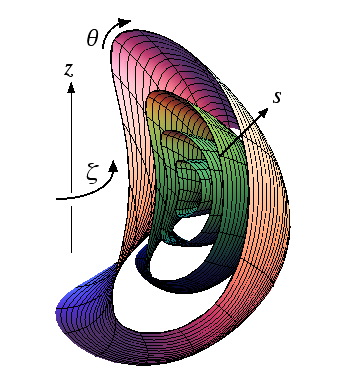}
	\caption{A sketch of the general curvilinear toroidal coordinate system described in the text.}
	\label{fig:Coordinates}
\end{figure}

In \Sec{Coords} we introduce our general, arbitrary background toroidal coordinate system $s,\theta,\zeta$, and an auxiliary poloidal angle $\Theta(s,\theta,\zeta)$ that allows us to define the quadratic flux in a form independent of the choice of $\theta$.
 In \Sec{Action} we introduce the magnetic action integral, its first and second variations and Hamilton's principle, while in \Sec{GhostQFMin} we introduce QFMin and (generalized) ghost-surface pseudo-orbits as alternative strategies for continuously deforming the action-minimax orbit associated  with an island chain into the corresponding action-minimizing orbit.

In \Sec{Comparison} we present numerical results for field-line Hamiltonians of the form $\chi_0(\psi) + \epsilon \chi_1(\psi,\theta,\zeta)$, where the flux function $\psi$ plays the role of a momentum canonically conjugate to $\theta$, and $\epsilon$ parametrizes the strength of the perturbation away from the integrable case described by the action-angle Hamiltonian $\chi_0$.
 Plots are presented comparing the uncorrected (i.e. with $\Theta = \theta$) QFMin and Lagrangian ghost curves of Ref.~\onlinecite{Hudson_Dewar_96},
 superposed on field-line puncture plots in a Poincar{\'e} surface of section.
 Two cases with different strengths of perturbation are shown, both quite strongly chaotic and both showing that the differences between even uncorrected QFMin and ghost curves are very small (except for some higher-order surfaces, in the more strongly chaotic case).
 This suggests that the two, seemingly very different, approaches to defining almost-invariant tori may be unified by appropriate choice of $\Theta$, and that this will differ from $\theta$ by an amount small in $\epsilon$.

In \Sec{ModHamPrin} we introduce a modified form of Hamilton's Principle that gives QFMin pseudo-orbits as extremizers of a pseudoaction.
 Section~\ref{sec:HamForm} gives the canonical, Hamiltonian form of this action principle, while \Sec{LagForm} discusses the transformation to the Lagrangian form.
 In \Sec{QFMinGhost} we derive a consistency condition that $\Theta$ must satisfy for corrected QFMin surfaces to be Lagrangian ghost surfaces, finding in \Sec{QFMinGhostPert} an expression for a choice of the auxiliary angle $\Theta$ that satisfies this criterion up to first order in $\epsilon$.
 The difference between uncorrected QFMin and ghost/corrected-QFMin pseudo-orbits is shown indeed to be very small, $O(\epsilon^2)$.

In \Sec{QFMinConstr} we sketch our finite-element variational method for numerical construction of QFMin surfaces using the new Hamilton's Principle introduced in \Sec{ModHamPrin}, and in \Sec{GhostConstr} we discuss the numerical construction of ghost surfaces via Galerkin projection onto the finite element basis.

Appendix~\ref{sec:QFMinEL1} contains a derivation of the Euler--Lagrange equation for QFMin pseudo-orbits in the canonical representation, and Appendix~\ref{sec:GenGhost} shows the relation between the generalized definition of ghost pseudo-orbit given in \Sec{GhostQFMin} and our more standard Lagrangian form \cite{Hudson_Dewar_96}, used in the numerical work and in \Sec{QFMinGhost}.

\section{Coordinates and fluxes}\label{sec:Coords}

 As depicted in \Fig{Coordinates}, we assume a general, essentially arbitrary curvilinear toroidal coordinate system $s(\vec{r}),\theta(\vec{r}),\zeta(\vec{r})$ has been established, where $\vec{r}$ is a point in Euclidean 3-space and $\theta$ and $\zeta$ are respectively poloidal and toroidal angles labeling points on the toroidal isosurfaces of $s$, nested around the curve along which $\theta$ is singular ($s$ increasing outward).
 We assume the nonorthogonal basis $\{\esup{s},\esup{\theta},\esup{\zeta}\} \equiv \{\grad{s},\grad{\theta},\grad{\zeta}\}$ is right handed, as is its reciprocal basis $\{\esub{s},\esub{\theta},\esub{\zeta}\} \equiv \{\dsub{s}\vec{r},\dsub{\theta}\vec{r},\dsub{\zeta}\vec{r}\}$.
 
The directed infinitesimal area element on an arbitrary surface $\Gamma$ is  $d\vec{S} \equiv d\theta d\zeta\,\vec{n}/\vec{n}\dotv\grad\theta\cross\grad\zeta$, where $\vec{n}$ is the unit normal at any point on $\Gamma$.
 Thus the net magnetic flux crossing an arbitrary torus $\Gamma$ (which we assume to contain the $\theta$-coordinate singularity curve) is
\begin{equation}
	\varphi_1[\Gamma] \equiv \int_0^{2\pi}\!\!\!\int_0^{2\pi}\!\!\!  d\theta d\zeta\,
	\frac{\vec{n}\dotv\vec{B}}{\vec{n}\dotv\grad\theta\cross\grad\zeta} \;.
	\label{eq:LFdef}
\end{equation}
This integral is independent of choice of coordinates.
 In fact the absence of magnetic monopoles implies that $\varphi_1$ vanishes identically, so it is independent of the choice of $\Gamma$ also, whether it be a magnetic surface (invariant torus of the field-line flow) or otherwise.

Thus,  to measure the amount by which $\Gamma$ departs from being a magnetic surface, we are led to define the positive definite \emph{quadratic flux} \cite{Dewar_Hudson_Price_94}, defined with the aid of a new generalized poloidal angle $\Theta (s,\theta,\zeta)$,
\begin{equation}
	\varphi_2[\Gamma] \equiv \frac{1}{2}\int_0^{2\pi}\!\!\!\int_0^{2\pi}\!\!\!  d\theta d\zeta\,
	\frac{\vec{n}\dotv\vec{B}}{\vec{n}\dotv\grad\theta\cross\grad\zeta}
	\frac{\vec{n}\dotv\vec{B}}{\vec{n}\dotv\grad\Theta\cross\grad\zeta} \;.
	\label{eq:QFdef}
\end{equation}
The quadratic flux $\varphi_2$ is independent of the choice of base coordinates $s,\theta,\zeta$, but depends on the choice of $\Theta$.

In the numerical work presented in this paper, $\Theta$ has been chosen equal to the given angle $\theta$.
 However in the formal development we distinguish it from $\theta$ so we can explore the consequences of making different choices, in particular whether it can be chosen so that QFMin tori coincide with ghost tori.

\section{Magnetic action integral}\label{sec:Action}

The field-line action  $\cal S$ \cite{Cary_Littlejohn_83} is a functional of a path ${\cal C}$ in Euclidean 3-space, points on which we take to be labeled by the toroidal angle $\zeta$, which thus takes on the role played by time in a more conventional Hamiltonian system.
In this paper we confine our attention to paths that are closed loops, with $\theta$ increasing by $2\pi p$ when $\zeta$ increases by $2\pi q$ ($p$ and $q > 0$ being mutually prime integers), so the average rate of increase of $\theta$ along the path is the rational fraction $p/q$, where the angular frequency $\iotabar$ is called the {\em rotational transform}.

The magnetic action is defined by
\be {\cal S}[{\cal C}] \equiv \int_{\cal C} {\bf A}\dotv\vec{dl} \equiv \int_0^{2\pi q} {\bf A}\dotv\dot{\vec{r}}\,d\zeta, \label{eq:magneticaction}
\ee
where the single-valued function $\vec{A}(\vec{r})$ is a magnetic vector potential for the magnetic field, $\vec{B}=\curl\vec{A}$, and $\vec{dl} \equiv \dot{\vec{r}}\,d\zeta$ is an infinitesimal line element tangential to  ${\cal C}$.
 A superscript dot denotes the total derivative with respect to $\zeta$, so that $\dot{\vec{r}}\dotv\grad\zeta \equiv 1$.
 \emph{Hamilton's Principle} is the statement that ${\cal S}$ is stationary, with respect to variations $\delta\vec{r}$ of ${\cal C}$, when ${\cal C}$ is a segment of a physical orbit (in our case a magnetic field line).
 If ${\cal C}$ is an open segment the variations are to be taken holding the endpoints fixed, but if (as we assume) ${\cal C}$ is a closed loop then the variations are unconstrained because the endpoint contributions cancel.
 Then, after integration by parts, we have the expansion for the total change in $\cal S$
\be \Delta {\cal S} = \int_0^{2\pi q} \left(\delta\vec{r}\dotv\frac{\delta\cal S}{\delta\vec{r}}
	+ \frac{1}{2}\delta\vec{r}\dotv\frac{\delta^2\cal S}{\delta\vec{r}\delta\vec{r}} \dotv\delta\vec{r} + \ldots\right)\, d\zeta , \label{eq:Svar}
\ee
where the first functional derivative is given by
\be \frac{\delta\cal S}{\delta\vec{r}} \equiv \esup{s}\frac{\delta {\cal S}}{\delta s} +  \esup{\theta}\frac{\delta {\cal S}}{\delta \theta} + \esup{\zeta}\frac{\delta {\cal S}}{\delta \zeta} = \dot{\vec r} \cross\vec{B} \;. \label{eq:Svar1}
\ee
Hamilton's Principle is now readily verified:  The Euler--Lagrange equation $\delta {\cal S}/\delta\vec{r} = 0$ is satisfied if $\dot{\vec{r}} = \vec{B}/B^{\zeta}$, i.e. on a magnetic field line.

The symmetrized Hessian operator is
\be 2\frac{\delta^2\cal S}{\delta\vec{r}\delta\vec{r}}
	& = & -\frac{d}{d\zeta}\vec{B}\cross\Id -\Id\cross\vec{B}\frac{d}{d\zeta}
	\nonumber\\ &&\mbox{}
	+ \dot{\vec{r}}\cross(\grad\vec{B})^{\rm T} - (\grad\vec{B})\cross\dot{\vec{r}}
	\; , \label{eq:Svar2}
\ee
where $\Id = \esub{s}\esup{s} + \esub{\theta}\esup{\theta} + \esub{\zeta}\esup{\zeta} = \esup{s}\esub{s} + \esup{\theta}\esub{\theta} + \esup{\zeta}\esub{\zeta}$ is the identity dyadic and superscript $^{\rm T}$ denotes the transpose.

Note also that variations $\delta\vec{r} = \vec{r}(\zeta + \delta\zeta) - \vec{r}(\zeta) = \vec{r} + \dot{\vec{r}}\delta\zeta + \frac{1}{2}\ddot{\vec{r}}(\delta\zeta)^2 + \ldots$ that simply relabel the path can be verified to leave $\cal S$ invariant for arbitrary $\delta\zeta(\zeta)$, as expected.
 Thus, to find a unique minimizer, we must suppress this degree of freedom in the allowed variations.
 To this end we constrain $\delta\vec{r}$ to the tangent plane of the poloidal surface of section $\zeta = \const$ at $\vec{r}$, denoting the constrained variation by $\Delta\vec{r} \equiv \esub{s} \delta s + \esub{\theta} \delta\theta$.
 (Provided $\vec{B}$, $\esub{s}$ and $\esub{\theta}$ are linearly independent, the two components of the Euler--Lagrange equation, $\delta {\cal S}/\delta s \equiv \esub{s}\dotv\dot{\vec r} \cross\vec{B} = 0$ and $\delta {\cal S}/\delta\theta \equiv \esub{\theta}\dotv\dot{\vec r} \cross\vec{B} = 0$ imply $\dot{\vec r} \cross\vec{B} = 0$, so the third component of the Euler--Lagrange equation, $\delta {\cal S}/\delta\zeta \equiv \esub{\zeta}\dotv\dot{\vec r} \cross\vec{B} = 0$, is redundant.)

\section {Ghost and QFMin surfaces}\label{sec:GhostQFMin}

\begin{figure}[htbp]
	\centering\includegraphics{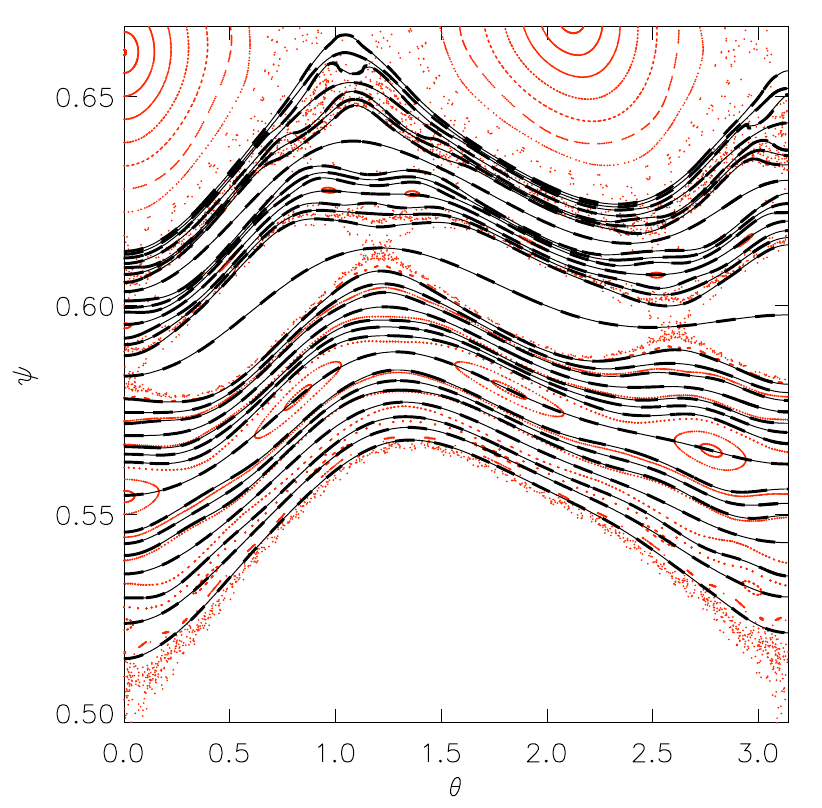}
	\caption{A comparison of the intersections of uncorrected QFMin surfaces (thick dashed lines) and Lagrangian ghost surfaces (thin lines) with the surface $\zeta = 0$, with the choices $s = \psi$, $\Theta = \theta$ described in Sec.~\ref{sec:HamForm}. The two almost-invariant surface definitions are almost indistinguishable in this moderately chaotic case, described in Sec.~\ref{sec:Comparison}. A \Poincare\ plot (red dots) is also shown.}
	\label{fig:comparisonlow}
\end{figure}

In an integrable system a continuous family of $(p,q)$-periodic orbits, each extremizing the action, exists, defining an invariant torus with rotational transform $p/q$.
 However such invariant tori are not structurally stable---small perturbations to the system destroy integrability  (see e.g. \Fig{comparisonlow}), leaving only isolated action-extremizing periodic orbits.
 In fact (assuming a twist condition holds), by the \Poincare--Birkhoff theorem \cite{Meiss_92}, only two distinct periodic orbits survive in a given $(p,q)$ island chain, namely the action-\emph{minimizing} orbit \cite{Min_vs_Max}, and an action-\emph{minimax} orbit.
 The minimizing orbit is a hyperbolically unstable ``X-point'' orbit in the chaotic separatrix region of the island chain while the minimax orbit threads the centers of the islands.
 The minimax orbit may be elliptically stable, or, after a period-doubling bifurcation, become unstable but continue as a $(p,q)$ hyperbolic orbit accompanied by a daughter $(2p,2q)$ elliptic orbit that does not affect the following discussion. 
 An almost-invariant torus, the ``ghostly remnant'' of an invariant torus, is formed from a family of pseudo-orbits, labeled by a parameter $\tau$, that ``fill in'' the regions between the minimizing and minimax orbits so as to form a torus $\Gamma$.

The pseudo-orbits of a \emph{ghost torus} are defined \cite{Hudson_Dewar_96} by deforming the minimax orbit via an action-gradient flow 
\be \frac{D\vec{r}}{D\tau} = -\frac{\delta\cal S}{\delta\vec{r}}\dotv\Proj_{\rm ghost}, \label{eq:gradflow}
\ee
where $D/D\tau$ denotes the total $\tau$-derivative at fixed $\zeta$, i.e. $D/D\tau \equiv (Ds/D\tau)\dsub{s} + (D\theta/D\tau)\dsub{\theta}$, and $\Proj_{\rm ghost}$ is a symmetric nonnegative dyadic.
 The most natural choice of $\Proj_{\rm ghost}$ might seem to be $\Proj_{\rm pol} \equiv \Id - \esup{\zeta}\esup{\zeta}/|\esup{\zeta}|^2$ projecting onto the poloidal tangent plane at $\vec{r}$, as this is independent of the choice of $s$ and $\theta$. However, as we show in Appendix~\ref{sec:GenGhost}, this does not correspond with that required to recover the usual Lagrangian definition \cite{Hudson_Dewar_96} of ghost surfaces.

At the periodic orbits, the action gradient is zero.
Beginning from the minimax orbit, we initially push the curve in the decreasing direction (provided by the eigenfunction of the Hessian with negative eigenvalue) and then evolve the curve according to the action gradient flow \Eqn{gradflow}.

Using  \Eqn{gradflow} in \Eqn{Svar} we find
\be \frac{D\cal S}{D\tau} = -\int_0^{2\pi q}\frac{\delta\cal S}{\delta\vec{r}}\dotv\Proj_{\rm ghost}\dotv\frac{\delta\cal S}{\delta\vec{r}} \, d\zeta , \label{eq:actionderiv}
\ee
so the sequence of pseudo-orbits tends monotonically toward the minimizing orbit, tracing out a surface, which we call the ghost surface.

\begin{figure}[htbp]
	\centering\includegraphics{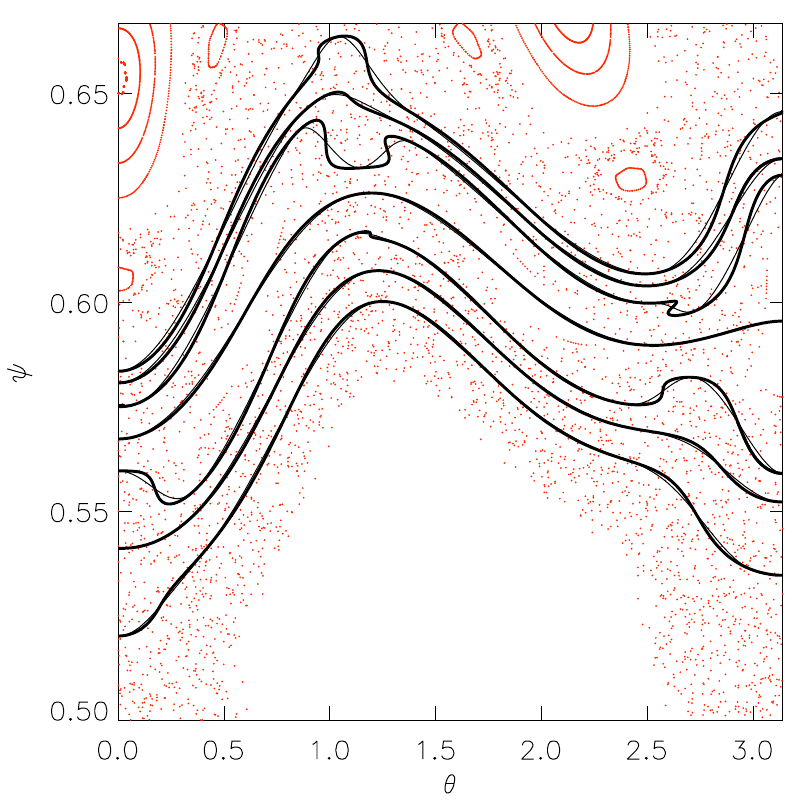}
	\caption{A comparison of uncorrected QFMin curves (thick lines) and ghost curves (thin lines) for a more strongly chaotic case described in the text. Some cases where QFMin curves violate the graph property are seen.}
	\label{fig:comparisonhigh}
\end{figure}

Ghost surfaces display several attractive properties (proved, for Lagrangian ghost orbits, in the case of symplectic maps \cite{Gole_01}; verified numerically for continuous time/magnetic field systems \cite{Hudson_Dewar_96}).
Their intersections with a surface of section are graphs when plotted in canonical phase-space coordinates, so that each line of constant $\theta$ crosses the ghost curve only once.
They are nonintersecting, and thus a discrete selection of ghost surfaces may be used as a framework for a generalized action-angle-like coordinate system for chaotic fields.
Furthermore, there appears to be a close correspondence between ghost surfaces and the isotherms resulting from strongly anisotropic heat-transport in chaotic magnetic fields \cite{Hudson_Breslau_08}.
However, as it stands, there is a significant disadvantage to their construction: the gradient flow vanishes as one approaches integrability.
In this limit, the construction of the ghost surfaces becomes arbitrarily slow!

A QFMin surface is one that minimizes $\varphi_2$ under deformations of $\Gamma$. 
In Ref.~\onlinecite{Dewar_Hudson_Price_94} 
(see also Appendix~\ref{sec:QFMinEL1}) it is shown that the Euler--Lagrange equation for this variational principle implies that $\Gamma$ is composed of pseudo-orbits tangential to the \emph{pseudo field}
\be \vec{B}_{\nu} \equiv \vec{B} - \nu \grad \Theta \cross \grad \zeta, \label{eq:pseudofield} \ee
where $\nu \equiv \vec{n}\dotv\vec{B}/\vec{n}\dotv\grad\Theta\cross\grad\zeta$ is constant on each such pseudo-orbit.
Intuitively, the pseudo field is constructed from the true field by adding a radial field that cancels the radial field caused by a perturbation away from a neighboring integrable field, while the poloidal field is unchanged \cite{Hudson_Dewar_98}.
Although QFMin curves in a Poincar{\'e} section are not guaranteed to have the graph property, QFMin pseudo-orbits are easier to construct than ghost orbits and thus it would be advantageous to find a QFMin formulation whose pseudo-orbits are also ghost orbits.

To construct a rational-rotational-transform QFMin surface we find periodic pseudo-orbits with rotational transform $p/q$.
We vary $\nu$ continuously over the range for which solutions can be found, the corresponding pseudo-orbits sweeping out ribbons that may be joined to form the entire surface $\Gamma$.
As the range includes $\nu = 0$, $\Gamma$ includes the closed field lines, both action-minimizing and minimax, associated with the magnetic island chain with the given rotational transform.

\section{Comparison}\label{sec:Comparison}

For illustration we use a model magnetic Hamiltonian (see Sec.~\ref{sec:HamForm}),  $\chi_0 + \chi_1$, consisting of an integrable part, \mbox{$\chi_0=\psi^2/2$}, and a perturbation,
\be \chi_1=\sum \chi_{m,n}(\psi)\cos(m\theta-n\zeta) \;. \ee
We use only two nonzero, $\psi$-independent, perturbation harmonics, \mbox{$\chi_{2,1}=0.0010$} and \mbox{$\chi_{3,2}=-0.0005$}, to drive islands at the
$\iotabar = 1/2$ and $\iotabar = 2/3$  rational surfaces.
The degree of chaos induced by the perturbations is illustrated by a \Poincare\ plot, shown with red dots in \Fig{comparisonlow}.
Quadratic-flux-minimizing surfaces (thick dashed lines)  constructed as in \Sec{QFMinConstr} and Lagrangian ghost surfaces (thin lines) constructed as in \Sec{GhostConstr} associated with $31$ rationals between these two islands are shown.
We term these QFMin surfaces \emph{uncorrected} because no attempt at transforming the coordinate system to improve agreement between QFMin and ghost surfaces has been made (i.e. we have taken $\Theta =\theta$).
 On the scale of the figure, the quadratic-flux minimizing surfaces and the ghost surfaces are indistinguishable.

As the degree of chaos increases, the quadratic-flux minimizing surfaces and the ghost surfaces appear to deviate, particularly those associated with the higher order rationals.
Increasing the perturbation amplitude to \mbox{$\chi_{2,1}=0.0020$} and \mbox{$\chi_{3,2}=-0.0010$}, the surfaces are shown in \Fig{comparisonhigh}.

The close agreement between QFMin curves and ghost curves in the case of moderate chaos for the coordinate choice used (action-angle coordinates in the unperturbed system) suggests a coordinate system in which they become identical may not be strongly perturbed away from action-angle coordinates.
 Below we point the way to achieving this unification.

\section{Modified Hamilton's Principle}\label{sec:ModHamPrin}

It has long been known \cite{Dewar_Khorev_95,Hudson_Dewar_96} that the quadratic flux $\varphi_2$ can be expressed in terms of action gradient, thus generalizing the QFMin concept to general Lagrangian/Hamiltonian systems and providing a link with the ghost surface approach.
 However, neither approach so far has provided a variational principle for individual pseudo-orbits.
In this section we present a new and very useful variational principle for the QFMin pseudo-orbits, modifying Hamilton's Principle by adjoining the constraint
\be {\cal A} \equiv \int_{\cal C} \Theta \grad\zeta \dotv\vec{dl} \equiv  \int_0^{2\pi q} \!\!\Theta\,d\zeta = \const. \label{eq:area}
\ee
This associates with a path  $\vec{r} = \vec{r}(\zeta)$ an ``area'' $\cal A$ under  the corresponding curve in the $(\zeta,\Theta)$ plane.  By writing $\Theta = (p/q)\zeta + {\Theta}_0 + \widetilde{\Theta}(\zeta)$, where $\widetilde{\Theta}$ is a periodic function averaging to $0$ and ${\Theta}_0$ is a constant approximately equal to the value of $\Theta$ at which the pseudo-orbit cuts the section $\zeta = 0$, we get ${\cal A}  = 2\pi q {\Theta}_0 + 2\pi^2 pq$.
Thus, constraining, $\cal A$ to be constant allows one to select a pseudo-orbit labelled by $\Theta_0$, and
\be \vec{r} = \vec{r}_{p,q}(\zeta|\Theta_0)\;, \quad\Theta \in [0,2\pi) \;,
\label{eq:rfamily} \ee
defines a family of paths covering a torus embedded in ${\Bbb R}^3$.

To implement the constraint \Eqn{area} using a Lagrange multiplier $\nu$ we define the \emph{pseudoaction}
\be {\cal S}_\nu \equiv {\cal S} - \nu {\cal A}, \label{eq:constrainedaction}
\ee
which is the same as the physical action \Eqn{magneticaction} with $\vec{A}$ replaced by the \emph{vector pseudopotential} $\vec{A}_{\nu} \equiv \vec{A} - \nu\Theta\grad\zeta$.
 Thus, the verification of Hamilton's Principle for QFMin pseudo-orbits goes through exactly as that in Sec.~\ref{sec:Action} for physical magnetic field lines (except that we must use the restricted variations, $\delta\vec{r} = \Delta\vec{r}$, so that the endpoint contribution $2\pi p\delta\vec{r}\dotv\grad\zeta$ vanishes): $\delta{\cal S}_{\nu}/\delta{s}  = \delta{\cal S}_{\nu}/\delta{\theta}  = 0$ implies $\dot{\vec r} \cross\vec{B}_{\nu} = 0$, where $\vec{B}_{\nu}$  is the pseudo magnetic field \Eqn{pseudofield}, as required for QFMin pseudo-orbits, for which $\dot{\vec r} = \vec{B}_{\nu}/B^{\zeta}$.

\section{Hamiltonian formulation}\label{sec:HamForm}

By exploiting gauge freedom we may write \mbox{${\bf A}=\psi\grad\theta-\chi(\psi,\theta,\zeta)\grad\zeta$}, for which ${\cal S}$ takes the familiar form
\be {\cal S}=\int_{\cal C}(\psi d\theta - \chi d\zeta) \;. \label{eq:canonS}
\ee
This is the canonical form for the vector potential, as the equations describing the field lines are \mbox{$\dot \theta = \partial \chi / \partial \psi$} and \mbox{$ \dot \psi = - \partial \chi / \partial \theta$}.
These are Hamilton's equations, with $\chi$ the magnetic-field-line Hamiltonian.
Henceforth we remove the arbitrariness in the radial coordinate $s$ by choosing it to be the flux function $\psi$, so that $s$ and $\theta$ are now canonically conjugate phase-space variables for the given magnetic field, and the return map generated by field lines intersecting the Poincar{\'e} section $\zeta = 0$ is area-preserving. 

From Hamilton's equations, we see that if $\chi$ depends only on $\psi$, e.g. \mbox{$\chi=\chi_0(\psi)$}, then $\psi$ and $\theta$ are action-angle coordinates \cite{Goldstein_80}: $\psi$ is constant, and $\theta$ increases linearly with $\zeta$ according to $\dot \theta = \iotabar(\psi)$.
 If the transform is rational, $\dot\theta=p/q$ for integer $p$ and $q$, there exists a continuous family of periodic field lines.
Each periodic field line may be identified, for example, by the poloidal angle where the field line intersects the \Poincare\ section $\zeta=0$.
Together, the periodic field lines form a {\em rational surface}.
However, in a nonintegrable system no transformation to action-angle coordinates exists and the Hamiltonian has the more general form \mbox{$\chi=\chi_0(\psi)+\chi_1(\psi,\theta,\zeta)$}.
 No matter how small $\chi_1$ is, the continuous family of periodic field lines is destroyed, being replaced by isolated action-minimizing and minimax orbits in an island chain.

In canonical coordinates the QFMin pseudoaction, \Eqn{constrainedaction}, is, from \Eqn{canonS}, ${\cal S}_{\nu} = \oint (\psi\dot{\theta} - \chi - \nu\Theta)d\zeta$, up to $O(\Delta\vec{r})$,
\be \delta {\cal S_{\nu}} & = & \int_0^{2\pi q}\!\!\!\! d\zeta
	\left[\left(\dot{\theta} - \dsub{\psi}\chi - \nu\dsub{\psi}\Theta \right)\delta\psi \right.
	\\ \nonumber &&\phantom{\int_0^{2\pi q}\!\!\!\! d\zeta} \left. - \left(\dot{\psi} + \dsub{\theta}\chi
	+ \nu\dsub{\theta}\Theta\right)\delta\theta \right] 
	\;, \label{eq:canonicalSnuvar1}
\ee
Setting the first variation to zero, we find Hamilton's equations for QFMin pseudo-orbits
\begin{eqnarray}
	\dot{\theta} & = & \dsub{\psi}\chi_{\nu} \label{eq:pseudoHameq1} \\
	\dot{\psi} & = & -\dsub{\theta}\chi_{\nu} \;, \label{eq:pseudoHameq2}
\end{eqnarray}
where $\chi_{\nu} \equiv \chi +\nu\Theta$. That is, $\nu\Theta$ acts as a scalar pseudopotential that displaces pseudo-orbits poloidally, away from the periodic orbits found when $\nu=0$.
\section{Lagrangian formulation}\label{sec:LagForm}

An arbitrary trial curve, ${\cal C}$, requires both the ``position curve,'' $\theta(\zeta)$, and the ``momentum curve,'' $\psi(\zeta)$, to be specified.

It is advantageous to reduce the number of dependent variables by transforming from the Hamiltonian phase-space description to the Lagrangian configuration-space description in the standard way, eliminating the momentum $\psi$ in favor of the velocity $\dot{\theta}$ given by \Eqn{pseudoHameq1}.
 [We assume the ``twist condition,'' $\chi_0^{\prime\prime}(\psi) > 0$, to allow unique inversion to give $\psi = \Psi_{\nu}(\dot\theta,\theta,\zeta)$, so this transformation may not be possible for systems with nonmonotonic rotational transform.]

The velocity curve then defines the momentum curve, making ${\cal S}_\nu$ a functional of the position curve alone, while partially extremizing
\be {\cal S_{\nu}}  =  \int_0^{2\pi q}\!\!\!\! d\zeta\, L_{\nu}(\theta,\dot{\theta},\zeta)
	\;, \label{eq:LagSnu}
\ee
where the pseudo-Lagrangian $L_{\nu} \equiv \Psi_{\nu}\dot{\theta} - \chi_{\nu}(\Psi_{\nu},\theta,\zeta)$. (The physical field-line Lagrangian $L$ and action ${\cal S}$ are obtained as the special case $\nu = 0$.)

Now the total perturbation of the pseudoaction is of the form
\be \delta {\cal S_{\nu}}  =  \int_0^{2\pi q}\!\!
	\delta\theta \frac{\delta{\cal S}_{\nu}}{\delta\theta}\, d\zeta
	\;, \label{eq:LagSnuvar}
\ee
where the Lagrangian pseudoaction gradient is
\be \frac{\delta{\cal S}_{\nu}}{\delta\theta}  =  \frac{\partial L_{\nu}}{\partial\theta}
	 - \frac{d}{d\zeta}\left(\frac{\partial L_{\nu}}{\partial\dot{\theta}}\right)
	\;. \label{eq:LagVar1}
\ee

\section{Reconciling QFMin and Lagrangian Ghost surfaces}\label{sec:QFMinGhost}

We now seek to choose $\Theta$ so that QFMin pseudo-orbits are also Lagrangian ghost pseudo-orbits as defined in  Appendix~\ref{sec:GenGhost}.
First note that, to reconcile \Eqn{pseudoHameq1} and \Eqn{Hameq1}, we require $\partial_{\psi}\Theta = 0$.
 That is,
\be \Theta \equiv \Theta(\theta,\zeta) \;.
\ee
Then $\Psi_{\nu} = \Psi$, $L_{\nu} = L - \nu\Theta$, $\partial L_{\nu}/\partial\dot{\theta} = \partial L/\partial\dot{\theta} $, and members of our family of QFMin pseudo-orbits
\be \theta = \theta(\zeta|\Theta_0) \;,
	\label{eq:thetafamily}
\ee
where $\Theta_0$ is an as yet arbitrary label [cf. \Eqn{rfamily}], satisfy the Euler--Lagrange equation
\be \frac{\delta{\cal S}_{\nu}}{\delta\theta}  = 
	\frac{\delta{\cal S}}{\delta\theta} - \nu(\Theta_0)\Theta_{\theta}(\theta,\zeta) = 0\;, 
	\label{eq:LagVar1recon}
\ee
with $\Theta_{\theta}(\theta,\zeta) \equiv \partial\Theta(\theta,\zeta)/\partial\theta$.

To reconcile  QFMin and ghost orbits we require that the family of pseudo-orbits defined by Eqs~(\ref{eq:thetafamily}) and (\ref{eq:LagVar1recon}) is the same family as is generated by \Eqn{Laggradflow}. Thus the labels $\Theta_0$ and $\tau$ must be functionally dependent: $\tau = \tau(\Theta_0)$, $d\tau = \tau'(\Theta_0)d\Theta_0$. Eliminating $\delta{\cal S}/\delta\theta$ between \Eqn{LagVar1recon} and \Eqn{Laggradflow} and observing that $D\Theta/D\Theta_0 \equiv (D\theta/D\Theta_0)\Theta_{\theta}$ we find the \emph{reconciliation condition}
\be \frac{D\Theta}{D\Theta_0}   = -\frac{\mu(\zeta)}{\tau'(\Theta_0)\nu(\Theta_0)}\left(\frac{D\theta}{D\Theta_0}\right)^2 \;.
	\label{eq:reconcilcon}
\ee
We now \emph{define} $\Theta_0$ so that, for all $\Theta_0$,
\be \tau'(\Theta_0)\nu(\Theta_0) & \equiv & -1 \;, \nonumber \\
	\theta(\zeta|\Theta_0+2\pi) & \equiv & \theta(\zeta|\Theta_0) + 2\pi \;,
\ee
choosing $\mu(\zeta)$ so that \Eqn{reconcilcon} satisfies the \emph{solvability condition} that the integral of both sides with respect to $\Theta_0$ over the interval $[0,2\pi]$ must be $2\pi$, giving
\be \mu(\zeta) = \left[\int_0^{2\pi}\frac{d\Theta_0}{2\pi}\left(\frac{\partial\theta(\zeta|\Theta_0)}{\partial\Theta_0}\right)^2\right]^{-1} \;.
\label{eq:solvability} \ee

\section{Perturbative construction of QFMin-ghost surfaces}\label{sec:QFMinGhostPert}

For example, to approach this task perturbatively, consider the Lagrangian of the class corresponding to the case studied numerically in \Sec{Comparison}
\be L = \frac{\dot{\theta^2}}{2} - \epsilon\!\!\!\!\sum^{\infty}_{m,n=-\infty}\!\!\!\! V_{m,n}\exp(im\theta-in\zeta) \;, \label{eq:modelL}
\ee
with the reality condition $V^{*}_{m,n} = V_{-m,-n}$ (superscript ${\cdot}^{*}$ denoting complex conjugation), and $\epsilon$ the expansion parameter.
 (The cases studied in \Sec{Comparison} thus correspond to the choices $V_{2,1} = \chi_{2,1}/2$, $V_{3,2} = \chi_{3,2}/2$, and $V_{m,n} = 0$ for $\{m,n\} \notin \{\{2,1\},\{-2,-1\},\{3,2\},\{-3,-2\}\}$.)

As the unperturbed system is integrable, the expansions of $\nu$, $\mu$, and $\Theta(\theta,\zeta)$ are of the form
\be \nu & = & \epsilon\nu_1 + \epsilon^2\nu_2 + \ldots \;, \nonumber\\
	\mu & = & \epsilon \mu_1 + \epsilon^2 \mu_2 + \ldots \;, \\
	\Theta & = & \theta + \sum_{m,n}\left(\epsilon \Theta^{(1)}_{m,n} + \epsilon^2 \Theta^{(2)}_{m,n} + \ldots\right)\exp i(m\theta - n\zeta) \nonumber \;,
	 \label{eq:epsExpansion}
\ee
and of the $(p,q)$ QFMin pseudo-orbits are of the form 
\be \theta(\zeta|\Theta_0) & = & \iotabar_{p,q}\zeta + \Theta_0 + \sum_{m,n}\left(\epsilon\theta^{(1)}_{m,n} + \epsilon^2\theta^{(2)}_{m,n} + \ldots\right)
	\nonumber\\
		& & \phantom{\iotabar_{p,q}\zeta + \Theta_0 + \sum_{m,n}} \times\exp i\left[\left(m\iotabar_{p,q}-n\right)\zeta + m\Theta_0\right] \;, \nonumber\\
	\psi(\zeta|\Theta_0) & = & \iotabar_{p,q}
	+ \sum_{m,n}i\left(m\iotabar_{p,q}-n\right)\left(\epsilon\theta^{(1)}_{m,n} + \epsilon^2\theta^{(2)}_{m,n} + \ldots\right)
	\nonumber\\
		& & \phantom{\iotabar_{p,q}\zeta + \Theta_0} \times\exp i\left[\left(m\iotabar_{p,q}-n\right)\zeta + m\Theta_0\right] \;,
	 \label{eq:OrbitepsExpansion}
\ee
where $\iotabar_{p,q} \equiv p/q$ and $\psi \equiv \partial L/\partial\dot{\theta} = \dot{\theta} = \theta_{\zeta}(\zeta|\Theta_0)$.

At $O(\epsilon)$ we find $\mu_1 = 0$ and
\be \nu_1(\Theta_0)  =  -\!\!\!\!\sum^{\infty}_{m,n=-\infty}\!\!\!\! im\delta_{mp,nq}V_{m,n}\exp(im\Theta_0)\;, \label{eq:nu1}
\ee
where the Kronecker delta $\delta_{mp,nq}$ selects Fourier coefficients such that $mp = nq$, resonant with $(p,q)$ pseudo-orbits.

The $O(\epsilon)$ contributions to the pseudo-orbit and pseudopotential are given by
\be \theta^{(1)}_{m,n}  = \Theta^{(1)}_{m,n}  =  \frac{im\bar{\delta}_{mp,nq}V_{m,n}}{(m\iotabar_{p,q} - n)^2}
\;, \label{eq:thetaTheta1}
\ee
where $\bar{\delta}_{mp,nq} \equiv 1 - \delta_{mp,nq}$ deletes the resonant components (which have been absorbed by $\nu_1$---unlike the KAM problem,  perturbation theory for almost-invariant tori is not inherently afflicted by small denominators, at least when $\iotabar$ is a low-order rational).

As the two Fourier coefficients in \Eqn{thetaTheta1} are the same, $\theta_1(\zeta|\Theta_0)  =  \Theta_1(\iotabar_{p,q}\zeta + \Theta_0,\zeta)$.
 However, $\Theta_1$ is \emph{not} used in the calculation of $\theta_1$: \emph{to first order, ghost and uncorrected QFMin pseudo-orbits are identical}, which, combined with  the vanishing of $\mu_1$, is consistent with the near indistinguishability of these two almost-invariant surfaces in \Fig{comparisonlow}.

At $O(\epsilon^2)$ we find
\be \theta^{(2)}_{m,n} & = & \bar{\delta}_{mp,nq}\sum_{m',n'}\!\!\!{}' \frac{im'(m+m')^2
	V_{m+m',n+n'}V_{m',n'}^{*}}{(m\iotabar_{p,q} - n)^2(m'\iotabar_{p,q} - n')^2} \nonumber\\
	&& \mbox{} - \frac{\bar{\delta}_{mp,nq}} {(m\iotabar_{p,q} - n)^2}\sum_{m',n'}\!\!{}'\, \nu^{(1)}_{m+m',n+n'} im'\Theta^{(1)*}_{m',n'}
\;, \label{eq:theta2}
\ee
where the prime on the sum over $m'$ and $n'$ indicates that the resonant terms, $m'p = n'q$, are to be deleted, and $\nu^{(1)}_{m,n} \equiv -imV_{m,n}\delta_{mp,nq}$.
 The term containing $\Theta^{(1)}$ gives the $O(\epsilon^2)$ difference between ghost pseudo-orbits and uncorrected QFMin pseudo-orbits.
 
We also see from \Eqn{solvability} that $\mu_2 =  - 2\epsilon^2\langle(\partial\theta_1/\partial\Theta_0)^2\rangle$, where $\langle\cdot\rangle$ denotes averaging over $\Theta_0$, so the correction factor $\mu$ in the definition of ghost surfaces required for reconciliation does deviate from unity by small amount, $O(\epsilon^2)$. In the numerical work presented in \Sec{GhostConstr} we take $\mu = 1$.

\section{Numerical construction of QFMin surfaces}\label{sec:QFMinConstr}

Previously \cite{Hudson_Dewar_98}, periodic pseudo field lines were found by integrating the Hamiltonian equations of motion Eqs~(\ref {eq:pseudoHameq1}--\ref {eq:pseudoHameq2}).
 The constrained action principle instead allows periodic pseudo orbits to be found variationally.
This is of great benefit for strongly chaotic fields, as a defining characteristic of chaos is the exponential separation of initially nearby trajectories at a rate given by the Lyapunov exponent.
Consequently, the intrinsic numerical error associated with field-line-following methods is guaranteed to increase as the trajectory becomes longer.
The method of Lagrangian integration, also called variational integration \cite{LMOW_04}, avoids these problems.
A Galerkin expansion of the trial curve,
\be \theta = \iotabar_{p,q}\zeta + \sum_{i=0}^{N-1} a_i u_i(\zeta)\;, \label{eq:thetaGalerkin}
\ee
in, say, $N$ continuous, $2\pi q$-periodic basis functions $u_i$, is inserted into the action principle and the unknown amplitudes $a_i$ are varied,
\be \delta\theta = \sum_{i=0}^{N-1} \delta a_i u_i(\zeta)\;, \label{eq:deltathetaGalerkin}
\ee
giving, from \Eqn{LagSnuvar},
\be \delta {\cal S_{\nu}}  & =  & \sum_{i=0}^{N-1}\left\langle u_i,\frac{\delta{\cal S}_{\nu}}{\delta\theta}\right\rangle a_i
	\nonumber \\ &&\!\!\!\!\!\!\!\!\!\!\mbox{}
	+ \frac{1}{2}\sum_{i=0}^{N-1}\sum_{j=0}^{N-1}
	a_i\left\langle u_i,\frac{\delta^2{\cal S}_{\nu}}{\delta\theta\delta\theta}u_j \right\rangle a_j +\dots
		\;, \label{eq:LagSnuvarGalerkin}
\ee
where the inner product $\langle f,g\rangle$, $f(\zeta)$ and $g(\zeta) $ arbitrary, is defined by
\be \langle f,g\rangle  =  \int_0^{2\pi q}\!\!\!\! d\zeta\, f g
	\;. \label{eq:InnerProd}
\ee

Extremizing curves are then obtained, for example, by finding a zero of the $N$-dimensional gradient $\langle \delta{\cal S}_{\nu}/\delta\theta,u_i \rangle$ using Newton's method.
This approach allows very high order periodic orbits to be found, even for quite strongly chaotic fields \cite{Hudson_06a}.

For numerical work, we must describe $\theta(\zeta)$ with a finite set of parameters.
It is simplest to use a piecewise linear description,
i.e. a finite-element expansion of $\theta(\zeta)$ in tent functions:
\be u_i(\zeta) & \equiv & \frac{1}{\Delta \zeta}\left[H(\zeta-\zeta_{i-1})H(\zeta_i-\zeta)(\zeta-\zeta_{i-1}) \right. \nonumber \\ 
	& & \left. \quad \mbox{} + H(\zeta-\zeta_i)H(\zeta_{i+1}-\zeta)(\zeta_{i+1}-\zeta)\right] 
	. \label{eq:Tent}
\ee
where $H(\cdot)$ is the Heaviside step function, $\zeta$ is to be evaluated mod $2\pi q$ to enforce periodicity [thus splitting the support of $u_0 \equiv u_N$ between $(0,\zeta_1)$ and $(\zeta_{N-1},\zeta_N)$], $\zeta_i \equiv i\Delta\zeta$, and $\Delta \zeta \equiv 2 \pi q /N$.
At first, the piecewise-linear approximation seems crude, but the great benefit is that the action integral can be calculated analytically in each interval \cite{Hudson_06a}.
The discretized pseudoaction ${\cal S}_{\nu}$ becomes a rapidly computable function of the $N$ independent parameters that describe the curve, \mbox{$\{\theta_{0},\theta_1,\dots,\theta_{N-1}\}$}, where $\theta_i \equiv (p/q)\zeta_i + a_i$.
 Extremal curves are found as zeros of the constrained action gradient, which are efficiently found using an $N+1$ dimensional Newton's method.
Note that the Hessian matrix $\langle u_i, (\delta^2{\cal S}_{\nu}/\delta\theta\delta\theta) u_j \rangle$ associated with the variations in the curve geometry is a cyclic tridiagonal matrix.

\section{Numerical construction of Lagrangian ghost surfaces}\label{sec:GhostConstr}

We discretize \Eqn{Laggradflow} using the Galerkin method. That is, we substitute the ansatz \Eqn{thetaGalerkin} into \Eqn{Laggradflow} and project onto the finite basis $\{u_i\}$,
\be	\sum_{i=0}^{N-1}\langle u_i,u_j \rangle\frac{D a_j}{D\tau} & = & -\left\langle u_i,\frac{\delta{\cal S}}{\delta\theta}\right\rangle
	\;. \label{eq:LaggradflowGalerkin}
\ee

For a basis with global support the matrix $\langle u_i,u_j \rangle$ is full and it must be inverted numerically to get $D a_j/D\tau$.
 For our finite-element basis the matrix is tridiagonal,
\be	\langle u_i,u_j \rangle  = \Delta \zeta\left[\delta_{i,j} +\frac{(\Delta\zeta)^2}{6}\frac{\delta_{i+1,j}-2\delta_{i,j}+\delta_{i-1,j}}{(\Delta\zeta)^2} \right]
	, \label{eq:FiniteEltOverlap}
\ee
where the Kronecker $\delta_{i,j}$ is the identity matrix.
 The term $(\delta_{i+1,j}-2\delta_{i,j}+\delta_{i-1,j})/(\Delta\zeta)^2$ is a finite-difference approximation to $d^2/d\zeta^2$, so, assuming the action gradient is a smooth function, the second term on the RHS is two orders in $\Delta\zeta$ smaller than the first and so can be neglected.
 Thus \Eqn{LaggradflowGalerkin} can be solved analytically to give
\be	\frac{D a_i}{D\tau} \equiv \frac{D \theta_i}{D\tau} = -\frac{1}{\Delta \zeta}\left\langle u_i,\frac{\delta{\cal S}}{\delta\theta}\right\rangle \equiv -\frac{1}{\Delta \zeta}\frac{\partial\cal S}{\partial\theta_i}
	\;. \label{eq:LaggradflowFinElt}
\ee

\section{Conclusion}\label{sec:Concl}

In this paper we have given generalized definitions of quadratic-flux-minimizing (QFMin) surfaces and ghost surfaces, formulating them in terms of the magnetic field-line action.
 We have gone further by introducing a new constrained Hamilton's Principle for QFMin pseudo-orbits that a) facilitates a reconciliation between QFMin surfaces and ghost surfaces, and b) provides a better algorithm than o.d.e. integration for calculation of QFMin orbits in strongly chaotic regions. 

QFMin and ghost orbits have been computed for a model magnetic field perturbed away from an integrable system described in action-angle coordinates, and found almost, but not quite, to coincide.
This is explained by finding a slight generalization of our previous ghost surface definition that allows QFMin and ghost surfaces to be fully reconciled, in principle, the change in the ghost surfaces being $O(\epsilon^2)$. Using perturbation theory we find a choice of poloidal angle that unifies QFMin and ghost surfaces up to second order in $\epsilon$, with the difference between ghost surfaces and uncorrected QFMin surfaces being $O(\epsilon^2)$.

It remains for future work to implement this reconciliation at all levels of nonlinearity and to explore whether it is beneficial to use further generalizations of the ghost surface definition.
 Also requiring further investigation is the question of why ghost surfaces coincide so closely with temperature isosurfaces \cite{Hudson_Breslau_08} for heat transport in chaotic magnetic fields.

The innovations in this paper should also be applicable in general $1\frac{1}{2}$-degree-of-freedom Hamiltonian systems and in the theory of area-preserving maps.

\appendix

\section{QFMin Euler--Lagrange equation}\label{sec:QFMinEL1}

Here we rederive in canonical coordinates the Euler--Lagrange equation for quadratic-flux-minimizing surfaces found by Dewar \emph{et al.} \cite{Dewar_Hudson_Price_94}.
Consider a toroidal magnetic field in canonical form, ${\bf B}=\grad \cross ( \psi \grad \theta - \chi \grad \zeta)$, with Jacobian $1/\sqrt g = \grad \psi \dotv \grad \theta \cross \grad \zeta$.
A toroidal surface is described by $\psi=P(\theta,\zeta)$.
We define the {\em tangential} dynamics using the equation \mbox{$\dot\theta=B^\theta / B^\zeta$} as a constraint, and require the ``radial'' dynamics to be confined to the surface, so that \mbox{$\dot\psi = P_\theta \dot\theta + P_\zeta$}, where $P_\alpha \equiv \partial_\alpha P$, $\alpha \in \{\theta,\zeta\}$.
The angle parametrization is arbitrary, and so let $\Theta=\Theta(\theta,\zeta)$ be a new poloidal angle.
We define $\nu$ as the projection, $\nu\equiv{\bf B}\cdot{\bf \bar N}$, of the magnetic field onto the vector \mbox{${\bf \bar N} \equiv ({\bf e}_\Theta + P_\theta {\bf e}_\psi)\cross ({\bf e}_\zeta + P_\zeta {\bf e}_\psi)$} normal to $\Gamma$, where here (and only here) ${\bf e}_\Theta$, ${\bf e}_\zeta$ are the derivatives of position with respect to $\Theta$, $\zeta$ at constant 
$\zeta$, $\Theta$ respectively: ${\bf e}_\Theta\equiv\partial{\bf x}/\partial\Theta|_\zeta$ and ${\bf e}_\zeta\equiv\partial{\bf x}/\partial\zeta|_\Theta$.
Combining these expressions, the tangential dynamics may then be written \mbox{$\dot \theta = \partial \chi / \partial \psi$} and \mbox{$\dot\psi=-\partial \chi / \partial \theta - \nu \Theta_\theta$}, which may be recognized as the pseudo field, \Eqn{pseudofield}.
The {\em generalized} quadratic-flux functional \Eqn{QFdef} can be written
\be \varphi_2 = \frac{1}{2}\int\!\!\int \! d\theta d\zeta\, \left({\bf B}\dotv {\bf N}\right) \,  \left({\bf B}\dotv {\bf \bar N}\right), \label{eq:genquadraticflux}
\ee
where \mbox{${\bf N} \equiv ({\bf e}_\theta + P_\theta {\bf e}_\psi)\cross ({\bf e}_\zeta + P_\zeta {\bf e}_\psi)$}, where ${\bf e}_\theta\equiv\partial{\bf x}/\partial\theta|_\zeta$, and now ${\bf e}_\zeta\equiv\partial{\bf x}/\partial\zeta|_\theta$.
The first variation in this functional due to variations in the surface is 
\be \delta \varphi_2 = \int\!\!\int \! d\theta d\zeta \delta P \sqrt g \left(B^\theta \partial_\theta + B^\zeta \partial_\zeta\right) \nu, \ee
where we have used $\partial(\sqrt g B^\beta)/\partial\alpha|_{\psi = P}=\partial_\psi(\sqrt g B^\beta)P_\alpha+\partial_\alpha(\sqrt g B^\beta)$, for $\alpha \in \{\theta,\zeta\}$ and $\beta \in \{\theta,\zeta\}$, to reflect the constraint that when $\theta$ or $\zeta$ vary, $\psi$ must also vary to remain on the surface.

\section{Generalized ghost surfaces}\label{sec:GenGhost}

In canonical coordinates \Eqn{Svar} becomes, to $O(\delta\vec{r})$,
\be \delta {\cal S} & = & \int_0^{2\pi q}\!\!\!\! d\zeta
	\left(\delta\psi\esub{\psi}+\delta\theta\esub{\theta} + \delta\zeta\esub{\zeta}\right)\dotv\frac{\delta\cal S}{\delta\vec{r}}
	\\ \nonumber & = & \int_0^{2\pi q}\!\!\!\! d\zeta
	\left[\left(\dot{\theta} - \dsub{\psi}\chi \right)\delta\psi - \left(\dot{\psi} + \dsub{\theta}\chi\right)\delta\theta
	+ \dot{\chi}\delta\zeta \right] 
	, \label{eq:canonicalSvar}
\ee
where the second form follows directly from \Eqn{canonS}. Identifying coefficients of $\delta\psi$ and $\delta\theta$ between the two forms,
\be 	\frac{\delta\cal S}{\delta\vec{r}} = 
	\left(\dot{\theta} - \dsub{\psi}\chi \right)\esup{\psi} - \left(\dot{\psi} + \dsub{\theta}\chi\right)\esup{\theta} 
	+ \dot{\chi}\esup{\zeta}
	\;. \label {eq:canonicalSvar1}
\ee
A slightly generalized definition of the Lagrangian definition \cite{Hudson_Dewar_96} of ghost surfaces can be found by choosing the projection operator in \Eqn{gradflow} to be 
\be 
	\Proj_{\rm ghost} = \frac{1}{\epsilon}\esub{\psi}\esub{\psi} + \frac{1}{\mu}\esub{\theta}\esub{\theta} \;,
	\label{eq:LagProj}
\ee
where $\epsilon \to 0$ is a switching factor required to obtain the Lagrangian description, as described below, and $\mu(\zeta) = O(1)$ is a factor we shall find necessary for unifying ghost and QFMin surfaces. Using \Eqn{LagProj} in \Eqn{gradflow}, the generalized action-gradient flow defining ghost surfaces in canonical coordinates becomes
\be \frac{D\psi}{D\tau} & = & -\frac{\dot{\theta}-\dsub{\psi}\chi}{\epsilon} \nonumber\\
	\frac{D\theta}{D\tau} & = & \frac{\dot{\psi} + \dsub{\theta}\chi}{\mu}
	\;. \label{eq:Hamgradflow}
\ee
Next we take the $\epsilon \to 0$ asymptotic limit, giving rise to two $\tau$-scales. On the short $\tau$-scale $\psi$ adjusts exponentially fast to enforce 
\be
	\dot{\theta} =\dsub{\psi}\chi \label{eq:Hameq1}
\ee 
on the long $\tau$-scale.

The action-gradient flow on the long $\tau$-scale defining ghost surfaces in Lagrangian form is thus
\be	\frac{D\theta}{D\tau} & = & -\frac{1}{\mu(\zeta)}\frac{\delta{\cal S}}{\delta\theta}
	\;, \label{eq:Laggradflow}
\ee
where ${\delta{\cal S}}/{\delta\theta} = -\dot{\psi} - \dsub{\theta}\chi \equiv \partial L/\partial\theta - (d/d\zeta)(\partial L/\partial\dot{\theta})$, with $L \equiv \psi\dot{\theta} - \chi$ and $\psi$ constrained to be a function of $\dot{\theta}$, $\theta$, and $\zeta$ by \Eqn{Hameq1}.

\section{Erratum}
The corrected version above includes the errata for Phys. Letts. A {\bf 373}, 4409--4415 (2009), http://dx.doi.org/10.1016/j.physleta.2009.10.005 \,:

On p. 4413, in the fifth line below Eq.~(23),  replace ``$d\tau = \tau'(\Theta_0)d\theta$'' with ``$d\tau = \tau'(\Theta_0)d\Theta_0$''.

On p. 4414, on the second line of Eq.~(32), replace
$$\mbox{}+\nu_1(\Theta_0)\frac{im\Theta^{(1)}_{m,n}}{(m\iotabar_{p,q} - n)^2}\;,$$
with
$$\mbox{}-\frac{\bar{\delta}_{mp,nq}} {(m\iotabar_{p,q} - n)^2}\sum_{m',n'}\!\!{}'\, \nu^{(1)}_{m+m',n+n'} im'\Theta^{(1)*}_{m',n'}\;,$$
and replace the sentence after this equation, 

``where the prime on the sum over $m'$ and $n'$ indicates that the resonant terms, $m'p = n'q$, are to be deleted.''

with

``where the prime on the sum over $m'$ and $n'$ indicates that the resonant terms, $m'p = n'q$, are to be deleted, and $\nu^{(1)}_{m,n} \equiv -imV_{m,n}\delta_{mp,nq}.$''

\section*{Acknowledgements}
Some of this work was supported by the Australian Research Council (ARC) and U.S. Department of Energy Contract No. DE-AC02-76CH03073 and Grant No. DE-FG02-99ER54546.
 We acknowledge a useful discussion with Prof. J.~D. Meiss on bifurcation of action-minimax orbits.

\bibliographystyle{elsarticle-num-names}
\bibliography{RLDBibDeskPapers} 

\end{document}